\documentclass[aps,showpacs,superscriptaddress]{revtex4}
\usepackage{epsfig}
\newcommand{\be}{\begin{equation}}
\newcommand{\ben}{\begin{eqnarray}}
\newcommand{\ee}{\end{equation}}
\newcommand{\een}{\end{eqnarray}}
\newcommand{\nnu}{\nonumber\\}
\newcommand{\bef}{\begin{figure}[htb]\centering}
\newcommand{\eef}{\end{figure}}
\newcommand{\sla}[1]{{#1}\!\!\!\slash}
\begin{document}
\title{Process dependent Sivers function and implication for single spin asymmetry in inclusive hadron production}

\date{\today}

\author{Leonard Gamberg}
\email{lpg10@psu.edu}
\affiliation{Division of Science, Penn State Berks, Reading, PA 19610, USA}

\author{Zhong-Bo Kang}
\email{zkang@bnl.gov}
\affiliation{RIKEN BNL Research Center,
                Brookhaven National Laboratory,
                Upton, NY 11973, USA}

\begin{abstract}
We study the single transverse spin asymmetries in the single inclusive particle production within the framework of the generalized parton model (GPM). By carefully analyzing the initial- and final-state interactions, we include the process-dependence of the Sivers functions into the GPM formalism. The modified GPM formalism has a close connection with the collinear twist-3 approach.
Within the new formalism, we make predictions for inclusive $\pi^0$ and direct photon productions at RHIC energies. We find the predictions are opposite to those in the conventional GPM approach.
\end{abstract}

\maketitle

%%%%%%%%%%%%
\section{Introduction}
\label{introduction}
Single transverse-spin asymmetries (SSAs) 
in both high energy lepton-hadron and 
hadronic scattering processes have attracted considerable attention 
from both experimental and theoretical communities 
over the  years \cite{D'Alesio:2007jt}. Generally, defined as
$A_N\equiv (\sigma(S_\perp)-\sigma(-S_\perp))/(\sigma(S_\perp)+\sigma(-S_\perp))$,
the ratio of the difference and the sum of the cross sections when the hadron's spin vector $S_\perp$ is flipped, 
SSAs have been consistently observed in various experiments at different collision energies \cite{HERMES, COMPASS, SSA-rhic}.

Much theoretical progress has been achieved in the recent years. 
An important realization is  the crucial role of the initial- and final-state interactions between the struck parton and the target remnant \cite{Brodsky:2002cx}, which provide the necessary phases that leads to the non-vanishing SSAs. These interactions can be accounted for by including the appropriate color gauge links in the gauge invariant transverse momentum dependent (TMD) parton distribution functions (PDFs) \cite{Collins:2002kn,TMD-dis,boermulders}. An important example is the quark Sivers function \cite{Siv90}, which represents the 
distribution of unpolarized quarks in a transversely polarized nucleon, through a correlation between the quark's transverse momentum and the nucleon polarization vector. 
They are believed to be (partially) responsible for the SSAs observed in the experiments.  

The details of the initial- and final-state interactions depend on the scattering process, thus the form of the gauge link in the Sivers function is process dependent \cite{mulders}. As a result, the Sivers function itself is non-universal. For example, it is the difference between the final-state interactions (FSIs) in semi-inclusive deep inelastic scattering (SIDIS) and the initial-state interactions (ISIs) in Drell-Yan (DY) process in $pp$ collision that leads to an opposite sign in the Sivers function probed in these two processes \cite{Collins:2002kn, boermulders, Kang:2009bp}. For hadron production in $pp$ collisions, typically the Sivers function has a more complicated relation relative
to those probed in SIDIS and DY processes \cite{mulders}; that is,  
there are only FSIs (ISIs)  in the SIDIS (DY) process, while both ISIs
and FSIs exist for single inclusive particle production. 

The SSAs for inclusive single particle production in hadronic collisions are among the earliest processes studied in experiments, starting from the fixed-target experiments in 1980s \cite{SSA-fixed-tgt}. Recently the experiments at Relativistic Heavy Ion Collider (RHIC) have also measured the SSAs of inclusive hadron production in $pp$ collisions over a wide range of energies~\cite{SSA-rhic}. Theoretically a QCD collinear factorization formalism at next-to-leading-power (twist-3) has been developed and been used in the phenomenological studies \cite{Efremov,qiu-sterman,inclupi,applytwist}. Alternatively, a more phenomenological approach has also been formulated in the context of generalized parton model (GPM) \cite{Anselmino,Boglione:2007dm,Anselmino:2008uy}, with the inclusion of spin and transverse momentum effects. In this approach  TMD factorization is assumed as a reasonable starting point~\cite{Anselmino};
at the same time, the leading twist TMD distributions (Sivers functions) are assumed to be universal (process-independent), thus the same as those in SIDIS process \cite{oldsiv,newsiv}.

In this paper we formulate the SSAs in inclusive single particle production within the framework of the GPM approach. However, instead of using a process-independent Sivers function, we will carefully examine the initial- and final-state interaction effects, and determine the process-dependent Sivers function. Further we find one can shift the process-dependence of the Sivers function to the squared hard partonic scattering amplitude under one-gluon exchange approximation, and these modified hard parts are very similar in form as those in the twist-3 collinear approach~\cite{inclupi} in terms of Mandelstam variables $\hat s,\hat t, \hat u$ (as we will demonstrate). This suggests a close connection between this modified GPM formalism and the twist-3 approach. However, it is important to mention that Mandelstam variables $\hat s,\hat t, \hat u$ are themselves a function of partonic intrinsic transverse momentum in the GPM approach. We comment on these issues at the end of Section~\ref{prosiver}, where we also show the modified GPM formalism can reproduce the twist-3 collinear factorization formalism in the leading order expansion in intrinsic transverse momentum $k_T$ (for contributions coming from initial and 
final state interactions, where the latter is equivalent up to a prefactor).  The rest of the paper is organized as follows: In Sec.~\ref{prosiver}, we introduce the GPM approach, demonstrate how to formulate the ISI and FSI effects, and discuss the connection to the twist-3 collinear factorization approach. In Sec.~\ref{numerics}, we estimate the asymmetry for inclusive pion and direct photon production at RHIC energy, and compare our predictions with those from the conventional GPM approach. We conclude our paper in Sec.~\ref{sum}.

%%%%%%%%%%%%%%%%
\section{Initial- and final-state interactions in single inclusive particle production}
\label{prosiver}
In this section, we introduce the basic ideas and assumptions of the GPM approach. Then we discuss how to formulate the initial- and final-state interactions for single inclusive particle production. Within the same framework of GPM approach, we thus derive a new formalism for the SSAs of single inclusive particle production, with the process-dependence of the Sivers function taken into account.

%%%%%%%%%%%%%%%%
\subsection{Generalized Parton Model}
The generalized parton model was introduced by Feynman and collaborators~\cite{Field:1976ve} as a generalization of the usual collinear pQCD approach. It was adapted and used to describe the SSAs for inclusive particle production~\cite{Anselmino,Boglione:2007dm,Anselmino:2008uy}, which has had considerable 
phenomenological success \cite{Boglione:2007dm}. According to this approach, 
for the inclusive production of large $P_{hT}$ hadrons (or photons), $A^\uparrow(P_A)+B(P_B)\to h(P_h)+X$, the differential cross section can be written as
\ben
E_h\frac{d\sigma}{d^3 P_h}=\frac{\alpha_s^2}{S}\sum_{a,b,c}\int \frac{dx_a}{x_a}d^2k_{aT}
f_{a/A^\uparrow}(x_a, \vec{k}_{aT})\int \frac{dx_b}{x_b}d^2k_{bT}
f_{b/B}(x_b, k_{bT}^2) \int \frac{dz_c}{z_c^2} D_{h/c}(z_c)H^U_{ab\to c}(\hat s,\hat t,
\hat u)\delta(\hat s+\hat t+\hat u),
\label{main}
\een
where $S=(P_A+P_B)^2$, $f_{a/A^\uparrow}(x_a, \vec{k}_{aT})$ is the TMD parton distribution functions with $k_{aT}$ the intrinsic transverse momentum of parton $a$ with respect to the light-cone direction of hadron $A$, and $D_{h/c}(z_c)$ is the fragmentation function. Since we will only consider the SSAs generated from the parton distribution functions in this paper, we have neglected the $k_T$-dependence in the fragmentation function. 
$H^U_{ab\to c}(\hat s,\hat t, \hat u)$ is the hard part coefficients with $\hat s, \hat t, \hat u$ the usual partonic Mandelstam variables. Eq.~(\ref{main}) can also be used to describe direct photon production, in which one replaces the  fragmentation function  $D_{h/c}(z_c)$ by $\delta(z_c-1)$, and $\alpha_s^2$ by $\alpha_{em}\alpha_s$.

To clearly specify the kinematics, we consider the center-of-mass frame of the two initial hadrons, in which one has $P_A^\mu=\sqrt{S/2}\,\bar{n}^\mu$ and $P_B^\mu=\sqrt{S/2}\, n^\mu$, with $\bar{n}^\mu=[1^+, 0^-, 0_\perp]$ and $n^\mu=[0^+, 1^-, 0_\perp]$ in light-cone components. For future convenience we  also 
define the hadronic Mandelstam invariants, $T=(P_A-P_h)^2$ and $U=(P_B-P_h)^2$. Additionally, the momenta of the partons in the partonic process $a(p_a)+b(p_b)\to c(p_c)+d(p_d)$ can be written as
\ben
p_a^\mu=\left[x_a\sqrt{\frac{S}{2}},\, \frac{k_{aT}^2}{x_a \sqrt{2S}},\,  \vec{k}_{aT}\right],
\qquad
p_b^\mu=\left[\frac{k_{bT}^2}{x_b \sqrt{2S}},\, x_b\sqrt{\frac{S}{2}},\,  \vec{k}_{bT}\right],
\een
where the momentum of parton $c$ is related to the final hadron as: $p_c=P_h/z_c$.

To study the SSAs, the PDFs $f_{a/A^\uparrow}(x_a, \vec{k}_{aT})$ in the transversely polarized hadron $A$ can be expanded as \cite{Anselmino,Boglione:2007dm,Anselmino:2008uy,Bacchetta:2004jz}
\ben
f_{a/A^\uparrow}(x_a, \vec{k}_{aT})=f_{a/A}(x_a, k_{aT}^2)+f_{1T}^{\perp a}(x_a, k_{aT}^2)
\frac{\epsilon^{k_{aT}S_{A} n\bar{n} }}{M},
\een
where $S_A$ is the transverse polarization vector, $M$ is the mass of hadron $A$, $f_{a/A}(x_a, k_{aT}^2)$ is the spin-averaged PDFs, and $f_{1T}^{\perp a}(x_a, k_{aT}^2)$ is the Sivers functions. Thus in GPM approach, the spin-averaged differential cross section is given by Eq.~(\ref{main}) with $f_{a/A^\uparrow}(x_a, \vec{k}_{aT})$ replaced by $f_{a/A}(x_a, k_{aT}^2)$, while the spin-dependent cross section is given by
\ben
E_h\frac{d\Delta\sigma}{d^3 P_h}&=&\frac{\alpha_s^2}{S}\sum_{a,b,c}\int \frac{dx_a}{x_a}d^2k_{aT}
f_{1T}^{\perp a}(x_a, k_{aT}^2)
\frac{\epsilon^{k_{aT}S_{A} n\bar{n} }}{M}
\int \frac{dx_b}{x_b}d^2k_{bT}
f_{b/B}(x_b, k_{bT}^2) 
\nnu
&&\times
\int \frac{dz_c}{z_c^2} D_{h/c}(z_c)H^U_{ab\to c}(\hat s,\hat t,
\hat u)\delta(\hat s+\hat t+\hat u),
\label{spin}
\een
and the SSA is given by the ratio,
\ben
A_N\equiv \left.E_h\frac{d\Delta\sigma}{d^3 P_h}\right/E_h\frac{d\sigma}{d^3 P_h}.
\een

As stated in the introduction, there are two assumptions in the GPM approach: one is that the spin-averaged and spin-dependent differential cross sections can be factorized in terms of TMD PDFs as in Eqs.~(\ref{main}) and (\ref{spin}), and the other one is that the Sivers functions is assumed to be universal and equal to those in SIDIS process, $f_{1T}^{\perp a}(x_a, k_{aT}^2)=f_{1T}^{\perp a, \rm SIDIS}(x_a, k_{aT}^2)$. In this paper we continue to work within the framework of the GPM approach, in other words, we will assume the TMD factorization is a reasonable 
phenomenological starting point. However, at the same time, we will take into account the initial- and final-state interactions. 
Since both ISIs and FSIs contribute for single inclusive particle 
production, in principle the Sivers functions in inclusive particle production in hadronic collisions should be different from those probed in SIDIS process.
We thus need to carefully analyze these ISIs and FSIs  for all the partonic scattering processes relevant to single inclusive particle production to determine
the proper Sivers functions to be used in the formalism. In other words, this new formalism will be
\ben
E_h\frac{d\Delta\sigma}{d^3 P_h}&=&\frac{\alpha_s^2}{S}\sum_{a,b,c}\int \frac{dx_a}{x_a}d^2k_{aT}
f_{1T}^{\perp a, ab\to cd}(x_a, k_{aT}^2)
\frac{\epsilon^{k_{aT}S_{A} n\bar{n} }}{M}
\int \frac{dx_b}{x_b}d^2k_{bT}
f_{b/B}(x_b, k_{bT}^2) 
\nnu
&&\times
\int \frac{dz_c}{z_c^2} D_{h/c}(z_c)H^U_{ab\to c}(\hat s,\hat t,
\hat u)\delta(\hat s+\hat t+\hat u),
\een
in which a {\it process-dependent Sivers function} denoted as $f_{1T}^{\perp a, ab\to cd}(x_a, k_{aT}^2)$ is used
rather than that from SIDIS  $f_{1T}^{\perp a, \rm SIDIS}(x_a, k_{aT}^2)$ as in the conventional GPM approach.

%%%%%%%%%%%%%%%%
\subsection{Initial- and final-state interactions}
In this subsection, we will discuss how to formulate the initial- and final-state interactions. 
The crucial point is that the existence of the Sivers function in the polarized nucleon relies on the initial- and final-state interactions between the struck parton and the spectators from the polarized nucleon through the gluon exchange. Thus by analyzing these interactions, one can determine the process dependent Sivers function $f_{1T}^{\perp a, ab\to cd}(x_a, k_{aT}^2)$ to be used for the corresponding partonic scattering $ab\to cd$. We start with the classic examples: the final-state interaction in SIDIS, and the initial-state interaction for DY process. To the leading order (one-gluon exchange), they are shown in Fig.~\ref{class}.
\bef
\psfig{file=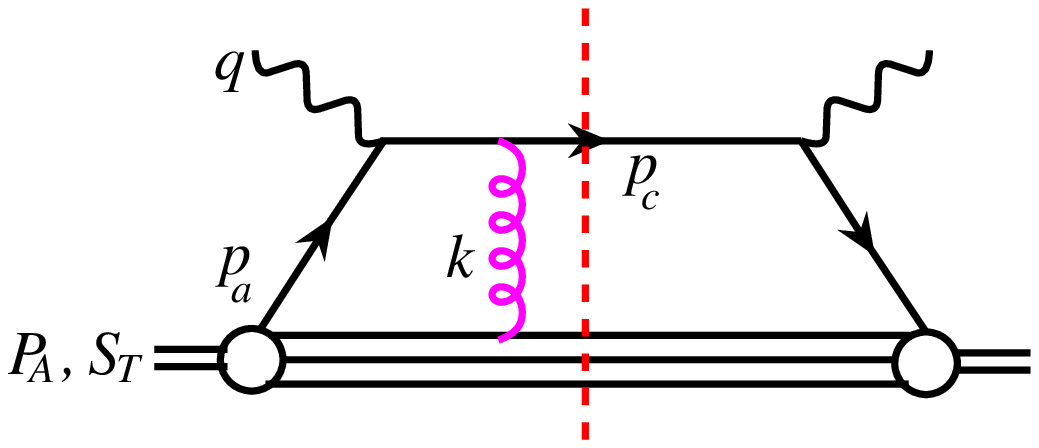, width=1.5in}
\hskip 0.3in
\psfig{file=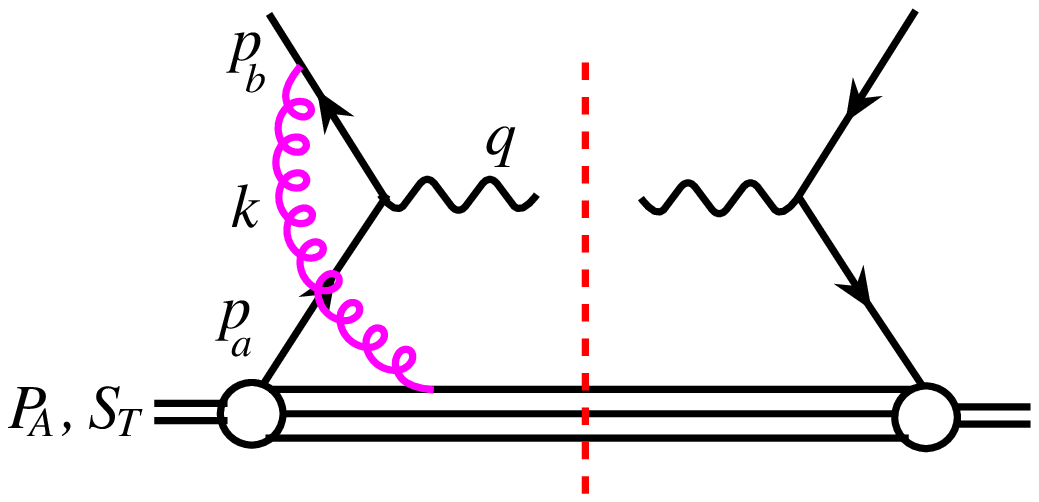, width=1.5in}
\caption{Final-state interaction in SIDIS (left) and initial-state interaction in DY (right) processes.}
\label{class}
\eef
For the SIDIS process $e(\ell)+p(P_A, S_T)\to e(\ell')+h+X$ with $Q^2=-q^2=-(\ell'-\ell)^2$, under the 
eikonal approximation, the final-state interaction (as in Fig.~\ref{class}(left)) leads to 
\ben
\bar{u}(p_c)(-ig)\gamma^-T^a\frac{i(\sla{p}_c-\sla{k})}{(p_c-k)^2+i\epsilon}\approx \bar{u}(p_c)\left[\frac{g}{-k^++i\epsilon}T^a\right],
\label{dis}
\een
where the gamma matrix $\gamma^-$ appears because of the interaction with a longitudinal polarized gluon ($\sim A^+$), and $a$ is the color index for this gluon.   The eikonal part (the term in the bracket) is  the first order contribution 
of the gauge link (in an expansion of the coupling $g$)
in the definition of a gauge-invariant TMD PDFs in SIDIS process, see Fig.~\ref{sidis_siv}(a). The imaginary part of the eikonal propagator $1/(-k^++i\epsilon)$ provides the necessary phase for the SSAs.
\bef
\psfig{file=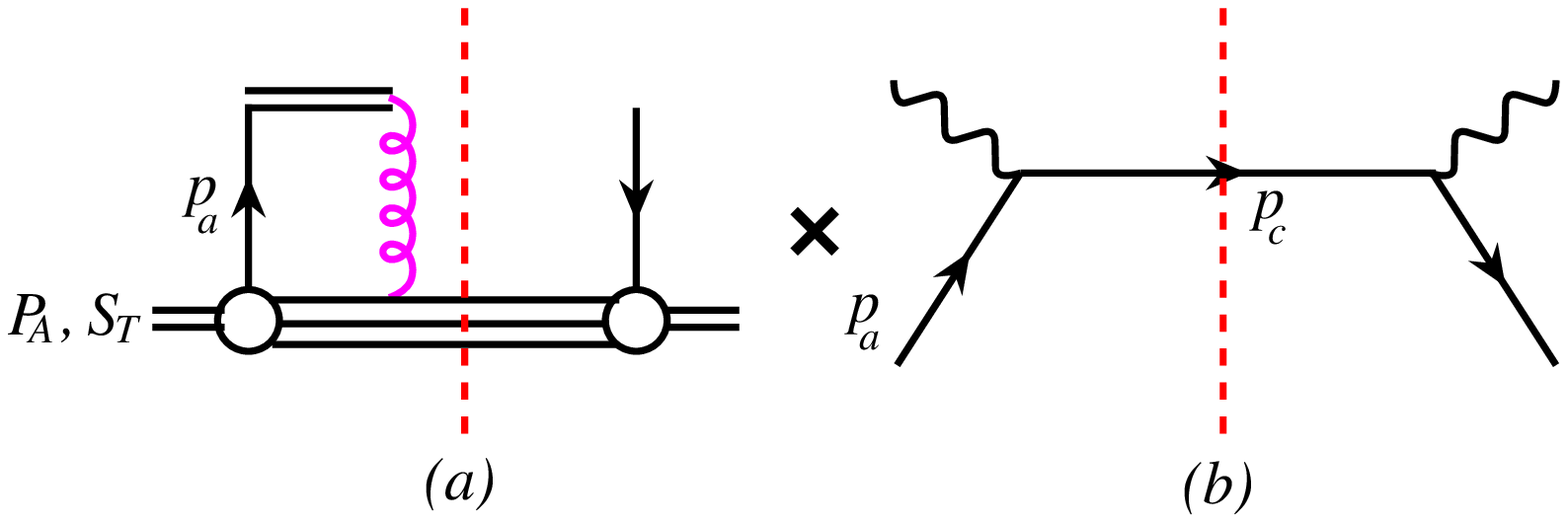, width=2.8in}
\caption{Sivers function in SIDIS process in the first non-trivial order (one-gluon exchange).}
\label{sidis_siv}
\eef

On the other hand, for DY process, the initial-state interaction (as in Fig.~\ref{class}(right)) leads to 
\ben
\bar{v}(p_b)(-ig)\gamma^-T^a\frac{-i(\sla{p}_b+\sla{k})}{(p_b+k)^2+i\epsilon}\approx
\bar{v}(p_b)\left[\frac{g}{-k^+-i\epsilon}T^a\right],
\een
which has the same real part and opposite imaginary part compared to SIDIS process. This leads to the fact that the spin-averaged TMD PDFs are the same, while the Sivers function will be opposite in SIDIS and DY processes. This conclusion can be generalized to all order, and has been proven to be true using parity and time-reversal invariant arguments \cite{Collins:2002kn,boermulders}.
\bef
\psfig{file=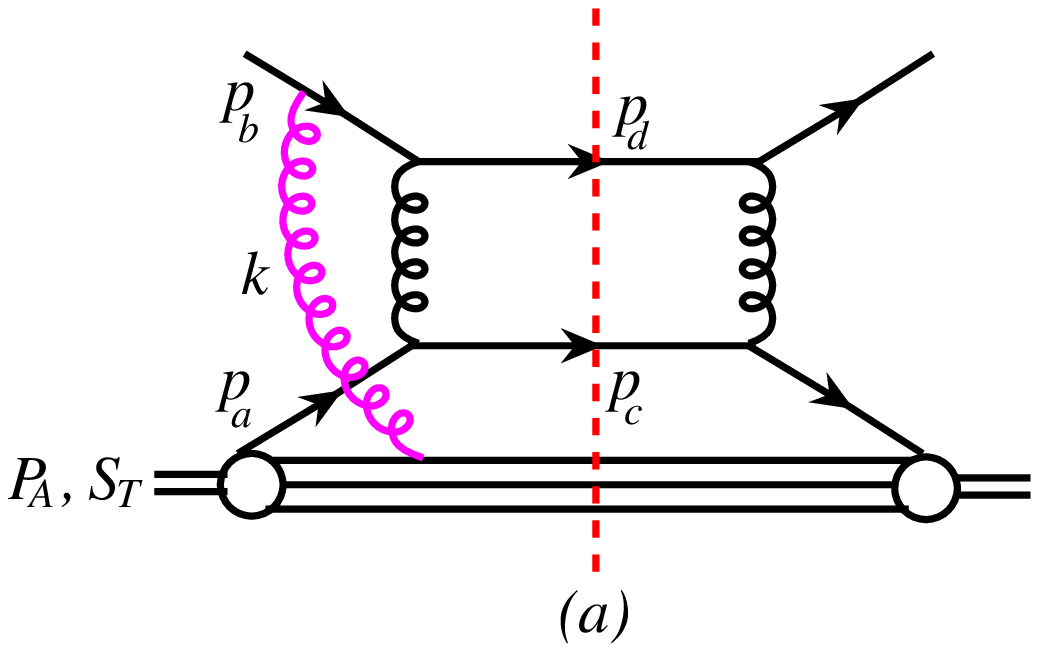, width=1.5in}
\hskip 0.15in
\psfig{file=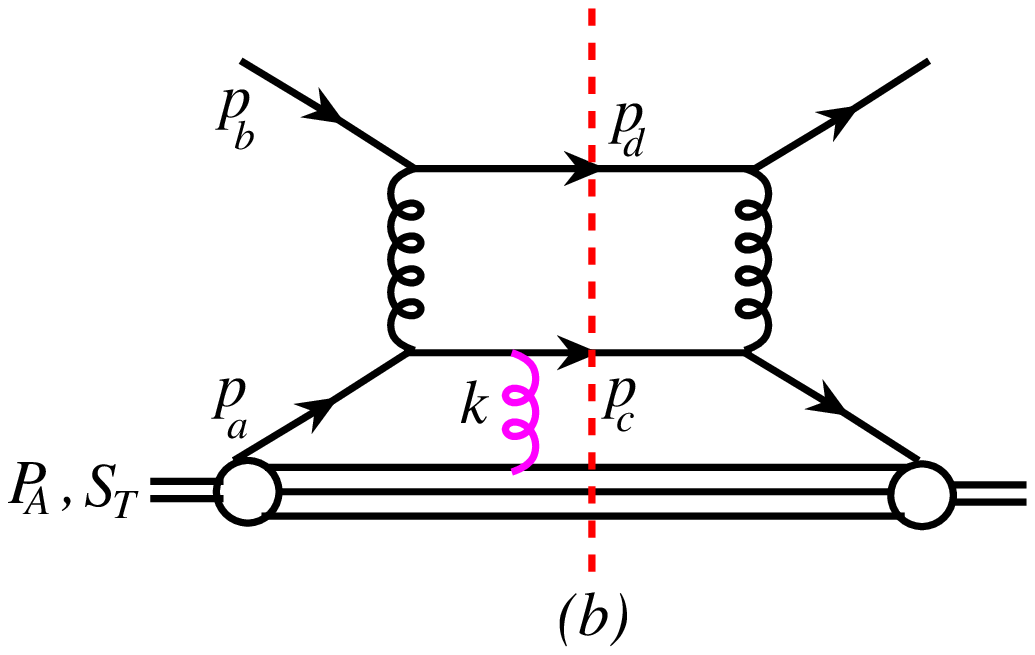, width=1.5in}
\hskip 0.15in
\psfig{file=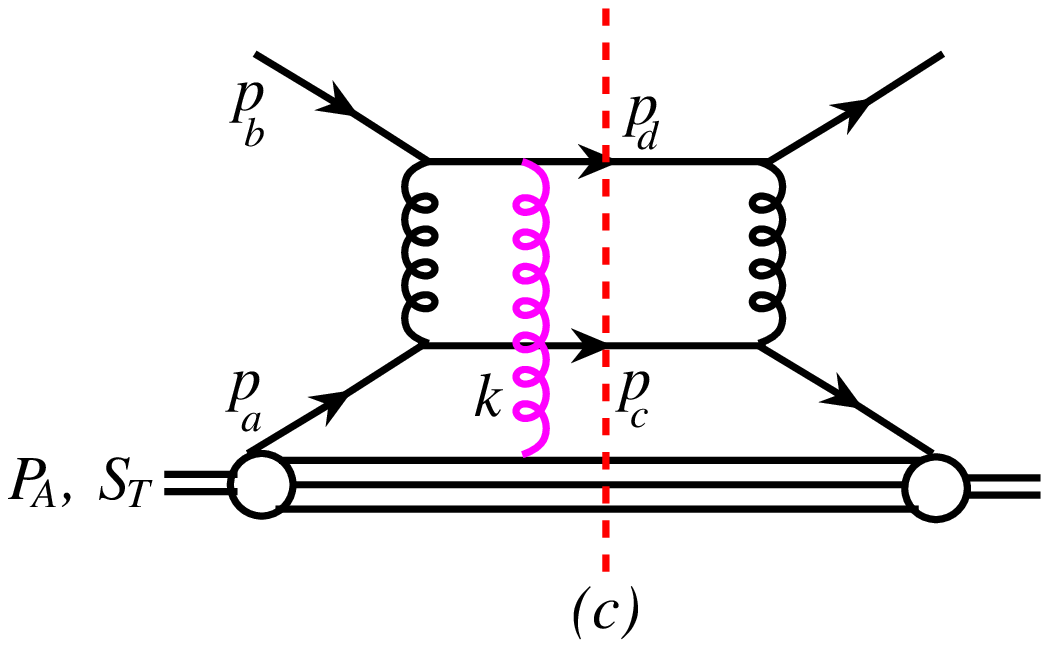, width=1.5in}
\hskip 0.15in
\psfig{file=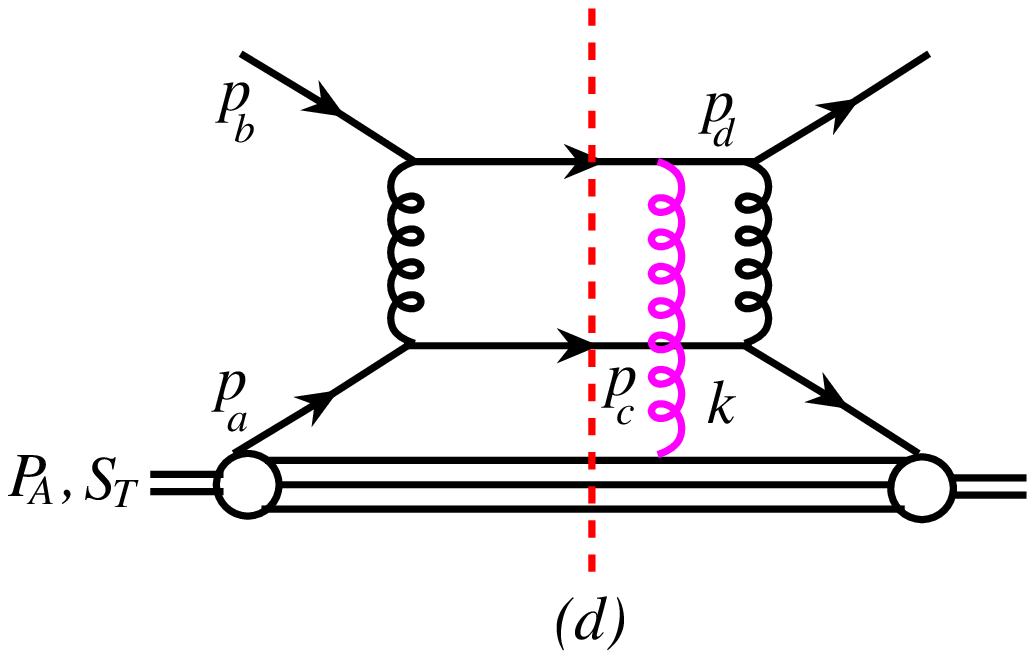, width=1.5in}
\caption{Initial- and final-state interactions in $qq'\to qq'$: (a) initial-state interaction, (b) final-state interaction, (c) and (d) the final-state interactions for the unobserved particle.}
\label{qqp}
\eef

Now let us turn to the case for inclusive single particle production in hadronic collisions, in which $2\to 2$ partonic scattering is the leading order contribution, where both initial- and final-state interactions contribute. 
We will start with a simple example: $qq'\to qq'$. 
Here the initial-quark $q$ is from the polarized nucleon, and  the  final-quark $q$ fragments to the final-state hadron. The one-gluon exchange approximation for the initial- and final-state interactions are shown in Fig.~\ref{qqp}. Under the eikonal approximation, for ISI Fig.~\ref{qqp}(a),
\ben
\frac{i(\sla{p}_b+\sla{k})}{(p_b+k)^2+i\epsilon}(-ig)\gamma^-T^a u(p_b)=\left[\frac{-g}{-k^+-i\epsilon}T^a\right] u(p_b).
\label{qqI}
\een
Likewise, for the FSI Fig.~\ref{qqp}(b), we have
\ben
\bar{u}(p_c)(-ig)\gamma^-T^a\frac{i(\sla{p}_c-\sla{k})}{(p_c-k)^2+i\epsilon}\approx \bar{u}(p_c)\left[\frac{g}{-k^++i\epsilon}T^a\right].
\label{qqF}
\een
Thus both interactions contribute to the phase $-i\pi \delta(k^+)$, which is the same as in the SIDIS process as in Eq.~(\ref{dis}). However, they will have different color flow. To extract the extra color factors for Fig.~\ref{qqp}(a) and (b) as compared to the usual $qq'\to qq'$ without gluon attachments, we resort to the method developed in \cite{qiu-sterman,inclupi,Qiu:2007ey}. We obtain the color factors $C_I$ ($C_{F_c}$) for initial (final)-state interaction
\ben
C_I=-\frac{1}{2N_c^2},
\qquad
C_{F_c}=-\frac{1}{4N_c^2},
\een
while the color factors for unpolarized cross section is given by
\ben
C_u=\frac{N_c^2-1}{4N_c^2}.
\een
In other words, the Sivers function in $qq'\to qq'$ should be the one as shown in Fig.~\ref{qqpsiv}, which comes from the sum of the ISIs and FSIs with the corresponding color factors $C_I$ and $C_{F_c}$ respectively.
\bef
\psfig{file=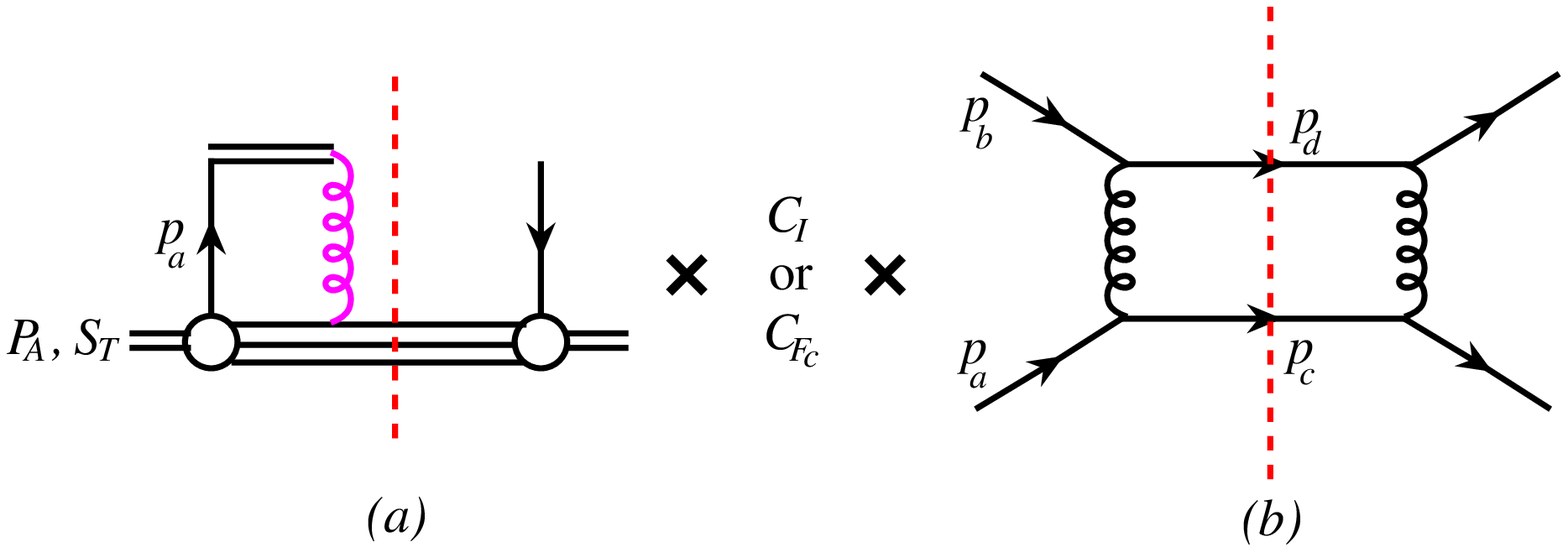, width=3.4in}
\caption{Sivers function in $qq'\to qq'$ from ISIs and FSIs, with the corresponding color factors $C_I$ and $C_{F_c}$ respectively.}
\label{qqpsiv}
\eef
Thus by comparing the imaginary part of the eikonal propagators in Eq.~(\ref{dis}) for SIDIS and those in Eqs.~(\ref{qqI}) and (\ref{qqF}) for 
ISI and FSI for $qq'\to qq'$, we immediately find the Sivers function probed in $qq'\to qq'$ process is related to those in SIDIS as follows
\ben
f_{1T}^{\perp a, qq'\to qq'}=\frac{C_I+C_{F_c}}{C_u}f_{1T}^{\perp a, \rm SIDIS}.
\een
Thus in the GPM model, using  the process dependent Sivers function, one should replace
\ben
f_{1T}^{\perp a, \rm SIDIS}H^U_{qq'\to qq'}\equiv f_{1T}^{\perp a, \rm SIDIS}\left[C_u h_{qq'\to qq'}\right],
\een
by the following form
\ben
f_{1T}^{\perp a, qq'\to qq'}H^U_{qq'\to qq'}=\frac{C_I+C_{F_c}}{C_u}f_{1T}^{\perp a, \rm SIDIS}H^U_{qq'\to qq'}=f_{1T}^{\perp a, \rm SIDIS}\left[C_I h_{qq'\to qq'} + C_{F_c}h_{qq'\to qq'}\right],
\een
where $h_{qq'\to qq'}$ is the partonic cross section without color factors included. For $qq'\to qq'$, one has
\ben
h_{qq'\to qq'}=2\frac{\hat s^2+\hat u^2}{\hat t^2}.
\een
Alternatively one can use 
$f_{1T}^{\perp a, \rm SIDIS}$ for the single inclusive particle production
while accounting for the process-dependence of the Sivers function, by
shifting the process-dependence to the hard parts. In other words, instead of using $H^U_{qq'\to qq'}$ in Eq.~(\ref{spin}) for the spin-dependent cross section, one should use
\ben
H^{\rm Inc}_{qq'\to qq'}\equiv H^{\rm Inc-I}_{qq'\to qq'}+H^{\rm Inc-F}_{qq'\to qq'},
\een
where 
\ben
H^{\rm Inc-I}_{qq'\to qq'}=C_I h_{qq'\to qq'},
\qquad
H^{\rm Inc-F}_{qq'\to qq'}=C_{F_c} h_{qq'\to qq'},
\een
are the corresponding hard parts related to initial- and final-state interactions, respectively.

There are many other partonic processes contributing to the single inclusive particle production. 
Similar to the analysis in $qq'\to qq'$, one needs to analyze each individual Feynman diagram accordingly, carefully moving the extra factors (process-dependence) from the corresponding Sivers function to the hard parts, thus obtaining $H^{\rm Inc-I}_{ab\to cd}$ and $H^{\rm Inc-F}_{ab\to cd}$ for every channel. The 
modified  formalism will be given in the next subsection.

There are some comments to our results presented to this point: 
in particular those displayed  in Fig.~\ref{qqpsiv}. It looks like Figs.~\ref{qqp}(a), (b) can be factorized into a convolution of Sivers function and a hard part function as shown in Fig.~\ref{qqpsiv}. However, this is not a TMD factorization in the strict sense. Currently TMD factorization theorems have been established for both SIDIS and DY processes \cite{Collins:1981uk, Ji:2004wu}. 
To the order we are studying, this means,
the one-gluon exchange diagram for SIDIS in Fig.~\ref{class} 
can be factorized into a convolution of a Sivers function $f_{1T}^{\perp a, \rm SIDIS}(x_a, k_{aT}^2)$ and a hard part function
$H(Q)$, as shown in Fig.~\ref{sidis_siv}. Here all the soft physics (those depending on $k_{aT}$) has been absorbed into
the Sivers function $f_{1T}^{\perp a, \rm SIDIS}(x_a, k_{aT}^2)$, and the hard part function $H(Q)$ only depends on the hard scale $Q$, not $k_{aT}$. On the other hand, for $qq'\to qq'$, we write the corresponding diagram Fig.~\ref{qqp}(a) into a similar form: a product of a Sivers function $f_{1T}^{\perp a, qq'\to qq'}(x_a, k_{aT}^2)$ and a hard part function $H_{qq'\to qq'}(\hat s$, $\hat t$, $\hat u)$, as shown in Fig.~\ref{qqpsiv}. But as we will comment later, besides the $k_{aT}$ dependence from the Sivers function, one will also need to keep the $k_{aT}$ dependence in the hard part functions $H_{qq'\to qq'}$, without which the SSAs will vanish in 
both the conventional GPM and this modified GPM  formalism. Even though this is not a TMD factorization, one hopes this formalism is a reasonable approximation. There are two reasons to suggest this might be the case. First of all, from phenomenological point of view, this formalism had some success \cite{Boglione:2007dm}. Secondly, as we will show in Section~\ref{prosiver} {\bf{D}} this formalism has a connection with the well-established collinear twist-3 approach \cite{inclupi}. In this respect, our identification of the color factors with the hard cross-sections is reminiscent of the results of the twist 3 approach (see in particular \cite{inclupi}). Indeed we will see that upon calculating all  partonic
processes that contribute from each channel, they have the same form in terms of Mandelstam variables $\hat s$, $\hat t$, $\hat u$, as compared to those in the twist-3 collinear factorization approach \cite{inclupi} (up to a prefactor associated with final state interactions).  

To close this subsection, we want to point out the following important fact: the interaction with the unobserved particle (the quark $q'$ for $qq'\to qq'$) vanishes after summing different cut diagrams \cite{qiu-sterman,inclupi,Yuan:2008vn}. To see this clearly, we have for Figs.~\ref{qqp}(c) and ~\ref{qqp}(d) 
\ben
\frac{1}{(p_d-k)^2+i\epsilon} \delta(p_d^2)\to-i\pi \delta((p_d-k)^2) \delta(p_d^2),\quad{\rm and}\quad
\frac{1}{p_d^2-i\epsilon} \delta((p_d-k)^2)\to+i\pi \delta((p_d-k)^2) \delta(p_d^2),
\een
respectively.  Since the remaining parts of the scattering amplitudes for these two diagrams are exactly the same except for the above pole contributions which are opposite to each other, the contribution from the unobserved
particle vanishes. This could also be used to explain why the inclusive DIS process, the SSA vanishes. As shown in Fig.~\ref{class} (left), we don't observe the final-state quark for the inclusive DIS process, thus the contribution from the cut to the left and to the right will  cancel which results in a vanishing asymmetry.

We want to emphasize that the above analysis holds true only under one-gluon exchange approximation. Going beyond one-gluon exchange, the Sivers functions are typically more complicated, there seems no simple relation (as extra color factors) to those in the SIDIS process \cite{tmdbreak}. 

\subsection{Single inclusive hadron production}
Now after carefully taking into account both initial- and final-state interactions, the more appropriate GPM formalism for spin-dependent cross section should be written as
\ben
E_h\frac{d\Delta\sigma}{d^3 P_h}&=&\frac{\alpha_s^2}{S}\sum_{a,b,c}\int \frac{dx_a}{x_a}d^2k_{aT}
f_{1T}^{\perp a, \rm SIDIS}(x_a, k_{aT}^2)
\frac{\epsilon^{k_{aT}S_{A} n\bar{n} }}{M}
\int \frac{dx_b}{x_b}d^2k_{bT}
f_{b/B}(x_b, k_{bT}^2) 
\nnu
&&\times
\int \frac{dz_c}{z_c^2} D_{h/c}(z_c)H^{\rm Inc}_{ab\to c}(\hat s,\hat t,
\hat u)\delta(\hat s+\hat t+\hat u),
\label{modified}
\een
where we have a new hard part function $H^{\rm Inc}_{ab\to c}$ instead of $H^U_{ab\to c}$ used in the conventional GPM approach. Here the process dependence in the Sivers function has been absorbed into $H^{\rm Inc}_{ab\to c}$, which can be written as
\ben
H^{\rm Inc}_{ab\to c}(\hat s,\hat t,\hat u)=H^{\rm Inc-I}_{ab\to c}(\hat s,\hat t,\hat u)+
H^{\rm Inc-F}_{ab\to c}(\hat s,\hat t,\hat u),
\label{mgpm}
\een
where $H^{\rm Inc-I}_{ab\to c}$ and $H^{\rm Inc-F}_{ab\to c}$ are associated with initial- and final-state interactions, respectively. The contributions  for the various contributing partonic subprocesses are 
given by
% q q' \to q q' and qbar q'bar \to qbar q'bar
\ben
H^{\rm Inc-I}_{qq'\to qq'}&=&-H^{\rm Inc-I}_{\bar q\bar q'\to \bar q\bar q'}=-\frac{1}{N_c^2}\left[\frac{\hat s^2+\hat u^2}{\hat t^2}\right]\, ,
\quad
H^{\rm Inc-F}_{qq'\to qq'}=-H^{\rm Inc-F}_{\bar q\bar q'\to \bar q\bar q'}=-\frac{1}{2N_c^2}\left[\frac{\hat s^2+\hat u^2}{\hat t^2}\right]
\\
% q q'bar \to q q'bar and qbar q' \to qbar q'
%\ben
H^{\rm Inc-I}_{q\bar q'\to q\bar q'}&=&-H^{\rm Inc-I}_{\bar qq'\to \bar q q'}=-\frac{N_c^2-2}{2N_c^2}\left[\frac{\hat s^2+\hat u^2}{\hat t^2}\right]\, ,
\quad
H^{\rm Inc-F}_{q\bar q'\to q\bar q'}=-H^{\rm Inc-F}_{\bar qq'\to \bar q q'}=-\frac{1}{2N_c^2}\left[\frac{\hat s^2+\hat u^2}{\hat t^2}\right]
%\een
% q q' \to q' q and qbar q'bar \to q'bar qbar
%\ben
\\
H^{\rm Inc-I}_{qq'\to q'q}&=&-H^{\rm Inc-I}_{\bar q\bar q'\to \bar q'\bar q}=-\frac{1}{N_c^2}\left[\frac{\hat s^2+\hat t^2}{\hat u^2}\right]\, ,
\quad
H^{\rm Inc-F}_{qq'\to q'q}=-H^{\rm Inc-F}_{\bar q\bar q'\to \bar q'\bar q}=\frac{N_c^2-2}{2N_c^2}\left[\frac{\hat s^2+\hat t^2}{\hat u^2}\right]
%\een
% q q'bar \to q'bar q and qbar q' \to q' qbar
%\ben
\\
H^{\rm Inc-I}_{q\bar q'\to \bar q' q}&=&-H^{\rm Inc-I}_{\bar qq'\to q'\bar q}=-\frac{N_c^2-2}{2N_c^2}\left[\frac{\hat s^2+\hat t^2}{\hat u^2}\right]\,  ,
\quad
H^{\rm Inc-F}_{q\bar q'\to \bar q' q}=-H^{\rm Inc-F}_{\bar qq'\to q'\bar q}=\frac{1}{N_c^2}\left[\frac{\hat s^2+\hat t^2}{\hat u^2}\right]
%\een
% q q \to q q and qbar qbar \to qbar qbar
%\ben
\\ \nonumber
H^{\rm Inc-I}_{qq\to qq}&=&-H^{\rm Inc-I}_{\bar q\bar q\to \bar q\bar q}=-\frac{1}{N_c^2}\left[\frac{\hat s^2+\hat u^2}{\hat t^2}
+\frac{\hat s^2+\hat t^2}{\hat u^2}\right]+\frac{N_c^2+1}{N_c^3}\frac{\hat s^2}{\hat t\hat u}\, ,
\\ 
H^{\rm Inc-F}_{qq\to qq}&=&-H^{\rm Inc-F}_{\bar q\bar q\to \bar q\bar q}=-\frac{1}{2N_c^2}\left[\frac{\hat s^2+\hat u^2}{\hat t^2}\right]
+\frac{N_c^2-2}{2N_c^2}\left[\frac{\hat s^2+\hat t^2}{\hat u^2}\right]+\frac{1}{N_c^3}\frac{\hat s^2}{\hat t\hat u}
%\een
% q qbar \to q' q'bar and qbar q \to q'bar q'
%\ben
\\ H^{\rm Inc-I}_{q\bar q\to q'\bar q'}&=&-H^{\rm Inc-I}_{\bar q q\to \bar q' q'}=\frac{1}{2N_c^2}\left[\frac{\hat t^2+\hat u^2}{\hat s^2}\right]\, ,
\quad
H^{\rm Inc-F}_{q\bar q\to q'\bar q'}=-H^{\rm Inc-F}_{\bar q q\to \bar q' q'}=\frac{N_c^2-2}{2N_c^2}\left[\frac{\hat t^2+\hat u^2}{\hat s^2}\right]
%\een
% q qbar \to q'bar q'
%\ben
\\ H^{\rm Inc-I}_{q\bar q\to \bar q' q'}&=&-H^{\rm Inc-I}_{\bar q q\to q'\bar q'}=\frac{1}{2N_c^2}\left[\frac{\hat t^2+\hat u^2}{\hat s^2}\right]\, ,
\quad
H^{\rm Inc-F}_{q\bar q\to \bar q' q'}=-H^{\rm Inc-F}_{\bar q q\to q'\bar q'}=\frac{1}{N_c^2}\left[\frac{\hat t^2+\hat u^2}{\hat s^2}\right]
%\een
% q qbar \to q qbar and qbar q \to qbar q
%\ben
\\ H^{\rm Inc-I}_{q\bar q\to q\bar q}&=&-H^{\rm Inc-I}_{\bar q q\to \bar q q}=-\frac{N_c^2-2}{2N_c^2}\left[\frac{\hat s^2+\hat u^2}{\hat t^2}\right]
+\frac{1}{2N_c^2}\left[\frac{\hat t^2+\hat u^2}{\hat s^2}\right]-\frac{1}{N_c^3}\frac{\hat u^2}{\hat s\hat t}\, , \nonumber
\\ 
H^{\rm Inc-F}_{q\bar q\to q\bar q}&=&-H^{\rm Inc-F}_{\bar q q\to \bar q q}=-\frac{1}{2N_c^2}\left[\frac{\hat s^2+\hat u^2}{\hat t^2}\right]
+\frac{N_c^2-2}{2N_c^2}\left[\frac{\hat t^2+\hat u^2}{\hat s^2}\right]+\frac{1}{N_c^3}\frac{\hat u^2}{\hat s\hat t}
%\een
% q qbar \to qbar q and qbar q \to q qbar
%\ben
\\ H^{\rm Inc-I}_{q\bar q\to \bar q q}&=&-H^{\rm Inc-I}_{\bar q q\to q\bar q}=-\frac{N_c^2-2}{2N_c^2}\left[\frac{\hat s^2+\hat t^2}{\hat u^2}\right]
+\frac{1}{2N_c^2}\left[\frac{\hat t^2+\hat u^2}{\hat s^2}\right]-\frac{1}{N_c^3}\frac{\hat t^2}{\hat s\hat u}\, , \nonumber
\\
H^{\rm Inc-F}_{q\bar q\to \bar q q}&=&-H^{\rm Inc-F}_{\bar q q\to q\bar q}=\frac{1}{N_c^2}\left[\frac{\hat s^2+\hat t^2}{\hat u^2}+\frac{\hat t^2+\hat u^2}{\hat s^2}\right]-\frac{N_c^2+1}{N_c^3}\frac{\hat t^2}{\hat s\hat u}
%\een
% q g \to q g and qbar g \to qbar g
%\ben
\\ H^{\rm Inc-I}_{qg\to qg}&=&-H^{\rm Inc-I}_{\bar qg\to \bar qg}=\frac{1}{2(N_c^2-1)}
\left[-\frac{\hat s}{\hat u}-\frac{\hat u}{\hat s}\right]+\frac{N_c^2}{2(N_c^2-1)}
\left[\frac{\hat s^2+\hat u^2}{\hat t^2}\frac{\hat u}{\hat s}\right]\, , \nonumber
\\
H^{\rm Inc-F}_{qg\to qg}&=&-H^{\rm Inc-F}_{\bar qg\to \bar qg}=\frac{1}{2N_c^2(N_c^2-1)}
\left[-\frac{\hat s}{\hat u}-\frac{\hat u}{\hat s}\right]-\frac{1}{N_c^2-1}
\left[\frac{\hat s^2+\hat u^2}{\hat t^2}\right]\, , 
%\een
% q g \to g q and qbar g \to g qbar
%\ben
\\ H^{\rm Inc-I}_{qg\to gq}&=&-H^{\rm Inc-I}_{\bar qg\to g\bar q}=\frac{1}{2(N_c^2-1)}
\left[-\frac{\hat s}{\hat t}-\frac{\hat t}{\hat s}\right]+\frac{N_c^2}{2(N_c^2-1)}
\left[\frac{\hat s^2+\hat t^2}{\hat u^2}\frac{\hat t}{\hat s}\right]\, ,\nonumber
\\
H^{\rm Inc-F}_{qg\to gq}&=&-H^{\rm Inc-F}_{\bar q g\to g\bar q}=-\frac{1}{2(N_c^2-1)}
\left[-\frac{\hat s}{\hat t}-\frac{\hat t}{\hat s}\right]-\frac{N_c^2}{2(N_c^2-1)}
\left[\frac{\hat s^2+\hat t^2}{\hat u^2}\frac{\hat s}{\hat t}\right]
%\een
% q qbar \to g g and qbar q \to g g
%\ben
\\ H^{\rm Inc-I}_{q\bar q\to gg}&=&-H^{\rm Inc-I}_{\bar q q\to gg}=-\frac{1}{2N_c^3}\left[\frac{\hat u}{\hat t}+\frac{\hat t}{\hat u}\right]-\frac{1}{N_c}\left[\frac{\hat t^2+\hat u^2}{\hat s^2}\right]\, ,\nonumber
\\
H^{\rm Inc-F}_{q\bar q\to gg}&=&-H^{\rm Inc-F}_{\bar q q\to gg}=-\frac{1}{2N_c}\left[\frac{\hat u}{\hat t}+\frac{\hat t}{\hat u}\right]+\frac{N_c}{2}\left[\frac{\hat t^2+\hat u^2}{\hat s^2}\frac{\hat u}{\hat t}\right]
\een
We also calculate the corresponding hard part functions for direct photon production, and they are given by
\ben
 H^{\rm Inc}_{qg\to \gamma q}&=&-H^{\rm Inc}_{\bar qg\to \gamma \bar q}=-\frac{N_c}{N_c^2-1}
e_q^2\left[-\frac{\hat t}{\hat s}-\frac{\hat s}{\hat t}\right]\, ,\quad
H^{\rm Inc}_{q\bar q\to \gamma g}=-H^{\rm Inc}_{\bar q q\to \gamma g}=\frac{1}{N_c^2}e_q^2
\left[\frac{\hat t}{\hat u}+\frac{\hat u}{\hat t}\right] \, .
\een
Here  again we note that all these hard part functions have the same form in terms of Mandelstam variables $\hat s$, $\hat t$, $\hat u$, compared to those in the twist-3 collinear factorization approach \cite{inclupi}: $H^{\rm Inc-I}_{ab\to c}$ and $H^{\rm Inc-F}_{ab\to c}$ have the same functional form as the corresponding ones $H^{\rm twist\mbox{-}3-I}_{ab\to c}$ and $H^{\rm twist\mbox{-}3-F}_{ab\to c}$ (defined below) in the twist-3 collinear factorization formalism, respectively. However, there are two differences in the formalisms. 
First, in the twist-3 collinear approach, the hard part functions are given by
\ben
H^{\rm twist\mbox{-}3}_{ab\to c}(\hat s,\hat t,\hat u)=H^{\rm twist\mbox{-}3-I}_{ab\to c}(\hat s,\hat t,\hat u)
+H^{\rm twist\mbox{-}3-F}_{ab\to c}(\hat s,\hat t,\hat u)\left(1+\frac{\hat u}{\hat t}\right),
\label{twist-3}
\een
i.e., there is an extra factor $(1+\hat u/\hat t)$ accompanying the hard part functions $H^{\rm twist\mbox{-}3-F}_{ab\to c}$ associated with final state interactions. However, in our modified GPM formalism as in Eq.~(\ref{mgpm}), there is no such factor. This difference can be traced back to the eikonal approximation we are using, see, e.g., Eq.~(\ref{qqF}),  where we only keep the pole contribution $-k^+ + i\epsilon$ in the denominator under this approximation. However, there is an extra term linear in $k_\perp$ ($\propto p_c\cdot k_\perp$) which  exists in the twist-3 collinear factorization formalism. This  leads to the extra factor $(1+\hat u/\hat t)$ for the final-state interaction contribution (for details, see Ref.~\cite{inclupi}). Second, in the twist-3 collinear factorization approach, all the parton momenta are collinear to the corresponding hadrons, thus $\hat s$, $\hat t$, $\hat u$ does not depend on the parton intrinsic transverse momentum. On the other hand, in the GPM approach the parton momenta involve intrinsic transverse momentum, thus $\hat s$, $\hat t$, $\hat u$ all depend on the the parton transverse momentum, $k_{aT}$ and $k_{bT}$. In fact, because of the existence of the linear $k_{aT}$-dependence in $\epsilon^{k_{aT}S_{A} n\bar{n} }$, one has to keep another linear $k_{aT}$-dependence from the rest of the integrand in Eq.~(\ref{modified}), otherwise the integral over $d^2k_{aT}$ vanishes. In other words, it is the linear
in $k_{aT}$ term in the hard part functions $H^{\rm Inc}_{ab\to c}(\hat s, \hat t, \hat u)$ and $\delta(\hat{s}+\hat{t}+\hat{t})$ that contributes to the asymmetry. Even with these two differences, the similarities in terms of $\hat s$, $\hat t$, $\hat u$ suggest that there are close connections between our modified GPM formalism and the twist-3 collinear factorization approach. We explore this potential connection in the next subsection.

\subsection{Connection to the twist-3 collinear factorization formalism}
As pointed out in the last subsection, it is the linear in $k_{aT}$ dependence from the rest of the integral in Eq.~(\ref{modified}) that contributes to the asymmetry. We thus make an expansion and keep only the linear in $k_{aT}$ terms. We will show that the leading term in this expansion has a close connection to the twist-3 collinear factorization formalism.

We start by specifying the partonic kinematics. Keeping the linear in $k_{aT}$ terms and dropping all the $k_{bT}$-dependence we have
$p_a^\mu\approx x_a P_A^\mu+k_{aT}$ and 
$p_b^\mu\approx x_b P_B^\mu$, 
thus
\ben
\hat{s}\approx x_a x_b S,
\qquad
\hat{t}\approx \frac{x_a}{z_c} T -\frac{2P_{hT}\cdot k_{aT}}{z_c},
\qquad
\hat{u}=\frac{x_b}{z_c}U.
\een
Thus we can write the $\delta$-function as
\ben
\delta(\hat s+\hat t+\hat u)=\frac{1}{x_bS+T/z_c}\delta\left(x_a-x-\frac{2P_{hT}\cdot k_{aT}}{z_c x_b S+T}\right) \quad {\rm where,}\quad
x_a=x+\frac{2P_{hT}\cdot k_{aT}}{z_c x_b S+T},
\label{xaform}
\een
and $x=-x_bU/(z_c x_b S+T)$ is independent of $k_{aT}$. 
Now performing the integrate over $x_a$ in Eq.~(\ref{modified}) and using the $\delta$-function we get,
\ben
E_h\frac{d\Delta\sigma}{d^3 P_h}&=&\frac{\alpha_s^2}{S}\sum_{a,b,c}\int d^2k_{aT}
\frac{\epsilon^{k_{aT}S_{A} n\bar{n}}}{M}
\frac{1}{x_a}
f_{1T}^{\perp a, \rm SIDIS}(x_a, k_{aT}^2)
\int \frac{dx_b}{x_b}
f_{b/B}(x_b) 
\nnu
&&\times
\int \frac{dz_c}{z_c^2} D_{h/c}(z_c)H^{\rm Inc}_{ab\to c}(\hat s,\hat t,
\hat u)\left.\frac{1}{x_bS+T/z_c}\right|_{x_a=x+\frac{2P_{hT}\cdot k_{aT}}{z_c x_b S+T}}.
\label{before}
\een
After replacing $x_a$ as above, one has
\ben
\hat{s}=\tilde{s}-\frac{\tilde{s}}{\tilde{u}}2P_{hT}\cdot k_{aT}/z_c,
\qquad
\hat{t}=\tilde{t}+\frac{\tilde{s}}{\tilde{u}}2P_{hT}\cdot k_{aT}/z_c,
\qquad
\hat{u}=\tilde{u},
\label{stu}
\een
where $\tilde{s}=x x_b S$, $\tilde{t}=xT/z_c$, $\tilde{u}=x_b U/z_c$ and they are all independent of $k_{aT}$. Note $\hat{s}+\hat{t}+\hat{u}=0$ implies $\tilde{s}+\tilde{t}+\tilde{u}=0$.  Now besides the $\epsilon^{k_{aT}S_{A} n\bar{n}}$, the linear in $k_{aT}$ contributions
in Eq.~(\ref{before}) can come from, either (a) $x_a$-dependence in $f_{1T}^{\perp a, \rm SIDIS}(x_a, k_{aT}^2)$, or (b) the $\hat{s}$- and $\hat{t}$-dependence in $H^{\rm Inc}_{ab\to c}(\hat s,\hat t, \hat u)$. This is because $x_a$, $\hat s$, and $\hat t$ are the only terms in Eq.~(\ref{before}) which depend linearly in $k_{aT}$. We now make $k_{aT}$ expansion one by one. First for contribution (a), since
\ben
\frac{\partial x_a}{\partial k_{aT}^\alpha}=\frac{2P_{hT\alpha}}{z_c x_b S+T},
\een
to the linear term in $k_{aT}$, we have
\ben
E_h\frac{d\Delta\sigma^{(a)}}{d^3 P_h}&=&\frac{\alpha_s^2}{S}\sum_{a,b,c}\int d^2k_{aT}
\frac{\epsilon^{k_{aT}S_{A} n\bar{n}}}{M}
k_{aT}^\alpha \frac{2P_{hT\alpha}}{z_c x_b S+T}
\frac{d}{dx_a}\left[\frac{f_{1T}^{\perp a, \rm SIDIS}(x_a, k_{aT}^2)}{x_a}\right]_{x_a\to x}
\int \frac{dx_b}{x_b}
f_{b/B}(x_b) 
\nnu
&&\times
\int \frac{dz_c}{z_c^2} D_{h/c}(z_c)H^{\rm Inc}_{ab\to c}(\tilde s,\tilde t,
\tilde u)\frac{1}{x_bS+T/z_c},
\label{after}
\een
where we have dropped all $k_{aT}$ dependence in $H^{\rm Inc}_{ab\to c}$, thus replacing the 
$k_{aT}$-dependent $\hat s$, $\hat t$, $\hat u$ by the $k_{aT}$-independent $\tilde s$, $\tilde t$, $\tilde u$ in $H^{\rm Inc}_{ab\to c}$. Then using 
\ben
\int d^2 k_{aT} k_{aT}^\beta k_{aT}^\alpha f_{1T}^{\perp a, \rm SIDIS}(x_a, k_{aT}^2)=-\frac{1}{2} \int d^2 k_{aT} \,g^{\beta\alpha}\,|\vec{k}_{aT}|^2 f_{1T}^{\perp a, \rm SIDIS}(x_a, k_{aT}^2),
\een
and the relation between the Sivers function and the Efremov-Teryaev-Qiu-Sterman function $T_{a,F}(x, x)$ \cite{boermulders},
\ben
T_{a,F}(x,x)=-\frac{1}{M}\int d^2k_{aT} |\vec{k}_{aT}|^2f_{1T}^{\perp a, \rm SIDIS}(x, k_{aT}^2),
\een
one can rewrite Eq.~(\ref{after}) as
\be
E_h\frac{d\Delta\sigma^{(a)}}{d^3 P_h}\hspace{-0.1cm}=\hspace{-0.1cm}\frac{\alpha_s^2}{S}\sum_{a,b,c}\hspace{-0.05cm}
\int \frac{dz_c}{z_c^2} D_{h/c}(z_c) 
\frac{\epsilon^{P_{hT}S_A n \bar{n}}}{z_c \tilde{u}}
\frac{1}{x}\left[T_{a,F}(x, x)\hspace{-0.05cm}-\hspace{-0.05cm}x\frac{d}{dx}T_{a,F}(x, x)\right]
\int \frac{dx_b}{x_b}
f_{b/B}(x_b) 
H^{\rm Inc}_{ab\to c}(\tilde s,\tilde t,
\tilde u)\frac{1}{x_bS+T/z_c}.
\nnu
\label{aftera}
\ee
We observe that this {\em form} is  the same as that in the twist-3 collinear factorization approach. 
In particular, note that  there is no $k_{aT}$-dependence in the hard part functions $H^{\rm Inc}_{ab\to c}$.
The difference to the twist-3 collinear factorization formalism \cite{inclupi} (as mentioned above) is 
the extra factor $(1+\hat u/\hat t)$ accompanying the hard part functions associated with final-state interactions, 
see Eqs.~(\ref{mgpm}) and (\ref{twist-3}).

However, in our modified GPM formalism, we have another contribution from (b), due to the $k_{aT}$-dependence from $H^{\rm Inc}_{ab\to c}(\hat s, \hat t, \hat u)$ in Eq.~(\ref{before}). Let's now study this contribution (b). As is 
explicit in Eq.~(\ref{stu}) $\hat{u}$ is independent of $k_{aT}$  while both $\hat{s}$ and $\hat{t}$ depend on $k_{aT}$. Since $\hat s+\hat t+\hat u=0$, one could then set $\hat t=-\hat s-\hat u$ in $H^{\rm Inc}_{ab\to c}$ and then expand only $\hat s$ in  $k_{aT}$. That is,
\ben
\left.\frac{\partial}{\partial k_{aT}^\alpha}H^{\rm Inc}_{ab\to c}(\hat s, \hat t, \hat u)\right|_{k_{aT}\to 0}=
\left.\frac{\partial \hat s}{\partial k_{aT}^\alpha} \frac{\partial}{\partial \hat s}H^{\rm Inc}_{ab\to c}(\hat s, -\hat s-\hat u, \hat u)\right|_{k_{aT}\to 0}
=-\frac{2\tilde s}{\tilde u}\frac{P_{hT\alpha}}{z_c}\frac{\partial}{\partial \tilde s}
H^{\rm Inc}(\tilde s, -\tilde s-\tilde u, \tilde u).
\een
Then we have the contribution (b)
\be
E_h\frac{d\Delta\sigma^{(b)}}{d^3 P_h}=\frac{\alpha_s^2}{S}\sum_{a,b,c}
\int \frac{dz_c}{z_c^2} D_{h/c}(z_c) 
\frac{\epsilon^{P_{hT}S_A n \bar{n}}}{z_c \tilde{u}}
\frac{1}{x}T_{a,F}(x, x)
\int \frac{dx_b}{x_b}
f_{b/B}(x_b) 
\left[-\tilde{s}\frac{\partial}{\partial \tilde{s}}H^{\rm Inc}_{ab\to c}(\tilde s,-\tilde s-\tilde u,
\tilde u)\right]\frac{1}{x_bS+T/z_c}.
\label{afterb}
\ee
Thus to the leading order (linear in $k_{aT}$ terms), the spin-dependent cross section in our modified GPM formalism can be written as
\ben
E_h\frac{d\Delta\sigma}{d^3 P_h}=E_h\frac{d\Delta\sigma^{(a)}}{d^3 P_h}+E_h\frac{d\Delta\sigma^{(b)}}{d^3 P_h},
\een
with the contributions (a) and (b) given by Eqs.~(\ref{aftera}) and (\ref{afterb}), respectively. The term (a) {\it almost} reproduces the twist-3 collinear factorization formalism in Ref.~\cite{inclupi} modular the  extra factor $(1+\hat u/\hat t)$ associated with final state interactions, for which the origin of the difference is understood in last subsection. On the other hand, for the extra term (b), theoretically how to interpret this ``mismatch'' and why the term (b) does not appear in the usual twist-3 collinear factorization formalism deserves further investigation \cite{GK}. Here it is important to note, from
the phenomenological perspective, as already shown in \cite{inclupi}, the derivative of the correlation 
function $T_{a,F}(x, x)$ is the dominant contribution to the SSAs,  
thus we expect the term (b), which contains no derivative, 
to play a less important role in generating the SSAs compared with term (a). 
In other words, even though this 
modified GPM has an extra piece compared with the well-known 
twist-3 collinear factorization formalism, phenomenologically (numerically)  this formalism could give a good approximation to the SSAs. This remains to be confirmed \cite{GK} because there is still a difference in term (a) on the extra factor 
$(1+\hat u/\hat t)$ associated with the final state interactions between the twist-3 collinear factorization approach and our modified GPM formalism. If this were the case, it will provide further support to the modified GPM approach to the SSAs. 

To close this section, we want to emphasize that the contribution calculated in Ref.~\cite{inclupi} only comes from the so-called soft-gluon-pole (SGP) in the twist-3 collinear factorization approach. However, there are also contributions from so-called soft-fermon-pole (SFP) \cite{SFP}. Even though our modified GPM formalism might capture the main feature of SGP contributions, it seems unlikely to reproduce the SFP contributions. In this respect 
 the twist-3 formalism is  ``internally complete'' in the sense that the collinear factorization is expected to hold for this formalism \cite{Qiu:1990cu}. Finally, while TMD factorization is assumed in both GPM and our modified GPM formalisms, it is likely 
not to hold in these processes~\cite{tmdbreak}. However, the extent to which it is broken is not known numerically. 
Thus,  calculations within (modified) GPM formalisms should bear this in mind and thus be used with extra care.

\section{Numerical estimate of the SSAs}
\label{numerics}
In this section, we will estimate the SSAs for single inclusive hadron and direct photon production in $pp$ collisions at RHIC energy by using our modified GPM formalism in Eq.~(\ref{modified}). We will compare our results with those calculated from the conventional GPM formalism as in Eq.~(\ref{spin}).

To calculate the spin-averaged cross section, we use GRV98 LO parton distribution functions \cite{Gluck:1998xa} along with a Gaussian-type $k_T$-dependence \cite{newsiv, oldsiv}. The hard part functions for different partonic scattering channels are available in the literature \cite{inclupi,Owens:1986mp,Kang:2010vd}. For the spin-dependent cross section, we use the latest Sivers functions from \cite{newsiv} which are extracted from the recent SIDIS experiments. To consistently use this set of Sivers function, we will use DSS fragmentation function \cite{deFlorian:2007aj}.   For the numerical predictions below, we work in a frame in which the polarized hadron moves in the $+z$-direction, choosing 
$S_\perp, P_{h\perp}$ along $y$- and $x$-directions, respectively, where all the relevant distribution functions and fragmentation functions evaluated at the scale $P_{h\perp}$~\cite{Anselmino}.

\bef
\psfig{file=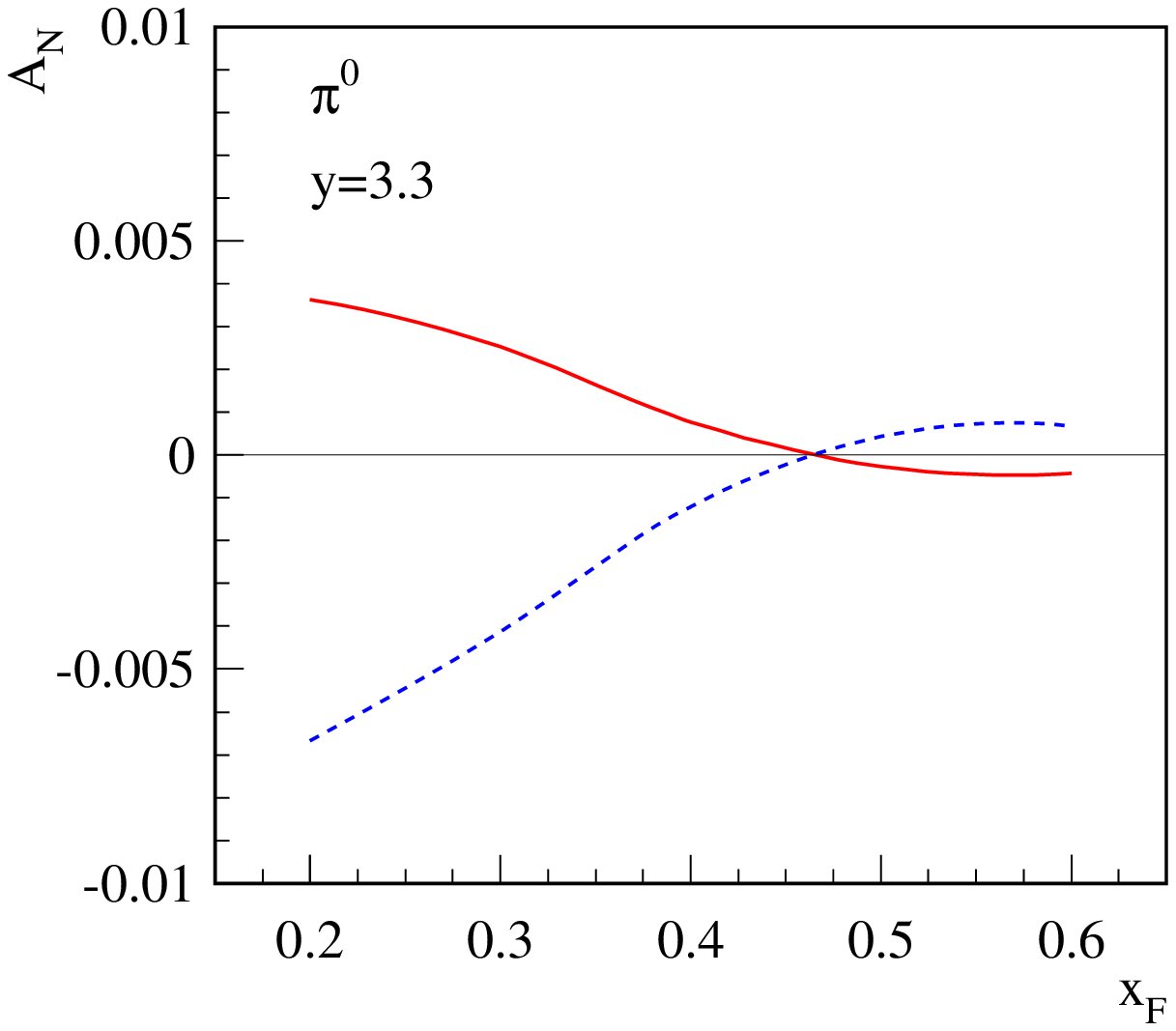, width=2.5in}
\hskip 0.2in
\psfig{file=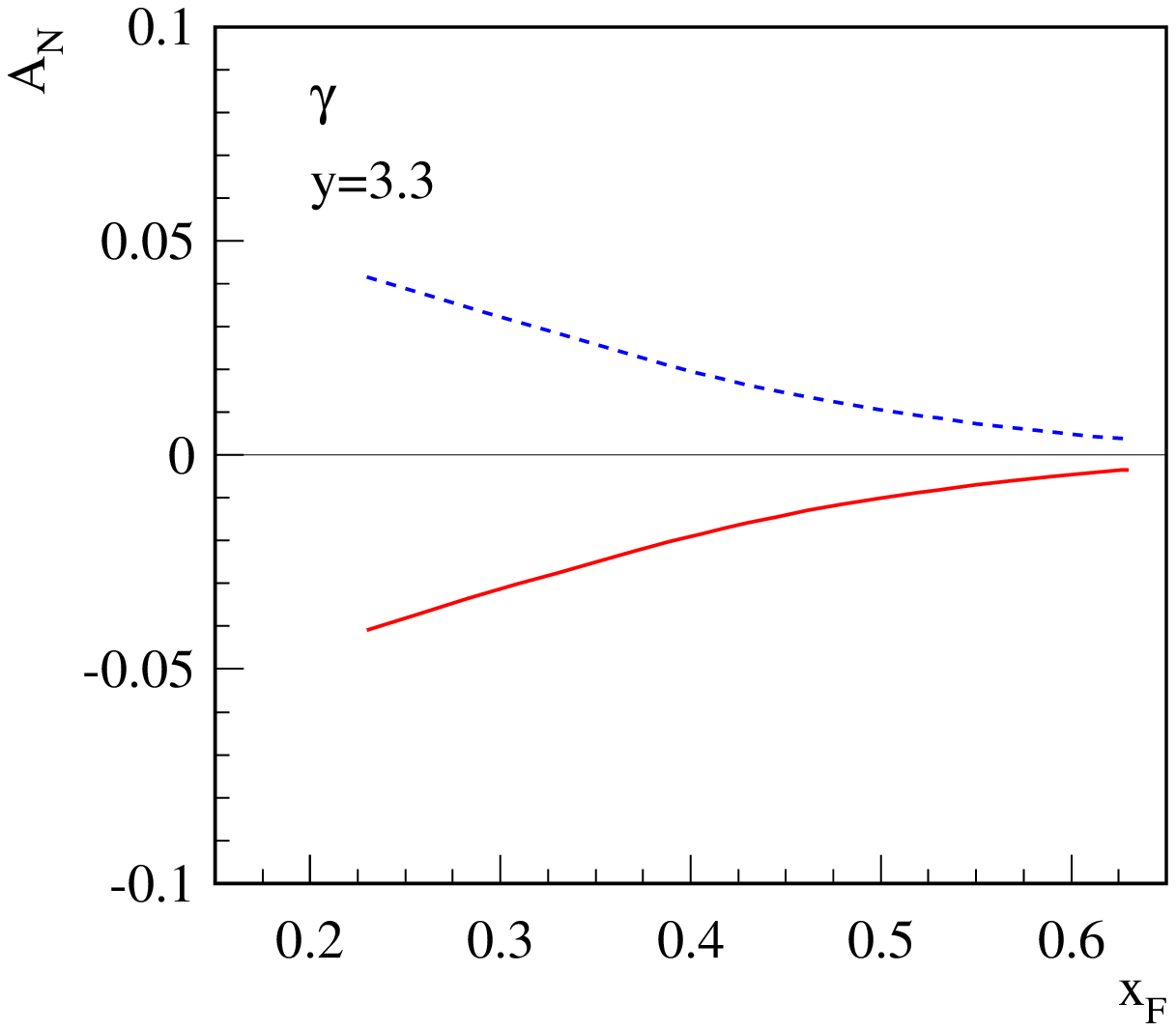, width=2.5in}
\caption{$A_N$ for inclusive particle production as a function of $x_F$ at RHIC energy $\sqrt{s}=200$ GeV: $p^\uparrow p\to\pi^0+X$ (left) and $p^\uparrow p\to\gamma+X$ (right). The dashed curves are for the conventional GPM calculation, and the solid curves are for our modified GPM calculation. We have used the latest Sivers function from \cite{newsiv}, and DSS fragmentation function \cite{deFlorian:2007aj}.}
\label{new}
\eef

In Fig.~\ref{new}, we plot the $A_N$ as a function of $x_F$ for inclusive $\pi^0$ (left) and direct photon (right) production at rapidity $y=3.3$ for RHIC energy $\sqrt{s}=200$ GeV.
The estimates using the conventional GPM formalism in Eq.~(\ref{spin}) are shown as dashed lines, while those using our modified GPM formalism in Eq.~(\ref{modified}) are shown as solid lines. One immediately see that for both inclusive $\pi^0$ and direct photon, $A_N$ change signs compare to the conventional GPM formalism. For $\pi^0$, the conventional GPM predicts a negative asymmetry (though very small from this set of Sivers functions), while the modified GPM formalism predicts a positive asymmetry. On the other hand, for direct photon, conventional GPM formalism predicts a positive asymmetry, while modified GPM formalism predicts that the asymmetry is negative, which is consistent with the predictions from twist-3 collinear factorization approach \cite{inclupi}. This can also be easily understood as follows. In the conventional GPM approach, one use $H^U$ in the calculation of the spin-dependent cross section. 
For direct photon production, the dominant channel comes from $qg\to \gamma q$, with \cite{inclupi,Owens:1986mp}
\be
H^U_{qg\to \gamma q}=\frac{1}{N_c}e_q^2
\left[-\frac{\hat t}{\hat s}-\frac{\hat s}{\hat t}\right]\, 
\ee
while  the hard part in the modified GPM formalism is  given by
\be
H^{\rm Inc}_{qg\to \gamma q}=-\frac{N_c}{N_c^2-1}
e_q^2\left[-\frac{\hat t}{\hat s}-\frac{\hat s}{\hat t}\right].
\ee
This introduces an extra color factor $-N_c^2/(N_c^2-1)$, thus opposite to the conventional GPM formalism. This prediction comes from the process-dependence of the Sivers functions, and has the same origin as in the photon+jet calculation \cite{Bacchetta:2007sz}. On the other hand, for the inclusive $\pi^0$ production, the dominant channel comes from $qg\to qg$, particularly in the forward direction, one has 
\ben
H^{\rm Inc}_{qg\to qg}=H^{\rm Inc-I}_{qg\to qg}+H^{\rm Inc-F}_{qg\to qg}
\to -\frac{N_c^2}{2(N_c^2-1)}\frac{2\hat s^2}{\hat t^2}-\frac{1}{N_c^2-1}\frac{2\hat s^2}{\hat t^2}=-\frac{N_c^2+2}{N_c^2-1}\frac{\hat s^2}{\hat t^2},
\een
where we have used that in the forward direction, $\hat t$ is small, while $\hat u\sim -\hat s$, whereas \cite{inclupi,Owens:1986mp}
\ben
H^U_{qg\to qg}=\frac{N_c^2-1}{2N_c^2}\left[-\frac{\hat s}{\hat u}-\frac{\hat u}{\hat s}\right]+\frac{\hat s^2+\hat u^2}{\hat t^2}\to \frac{2\hat s^2}{\hat t^2}.
\een
We thus also see the sign is reversed in our modified GPM formalism compared with the conventional GPM approach.

We observe that the $x_F$-dependence in both modified and conventional GPM formalisms are different from those observed in the RHIC experiments where larger asymmetries have  been observed in the forward direction (large $x_F$)~\cite{SSA-rhic}. Of course, in order to have a comparison with the experimental data for inclusive hadron production at RHIC experiments, one must include both Sivers (as studied in this paper) and Collins effects \cite{Collins:1992kk}. The latter describes a transversely polarized
quark jet fragmenting into an unpolarized hadron, whose
transverse momentum relative to the jet axis correlates
with the transverse polarization vector of the fragmenting quark. 
This latter correlation  can also generate the transverse spin asymmetry (which is not studied here). Currently  attempts at 
global fitting with both SIDIS and $pp$ experimental data are ongoing \cite{Anselmino:2008uy}. We encourage the use of the modified GPM formalism in such a global analysis, to study the effect of the
associated ISIs and FSIs (process-dependence of the Sivers functions).
We also emphasize~\cite{Bacchetta:2007sz}  that there is only Sivers contribution in direct photon production. Since the modified and conventional GPM predict opposite asymmetries, 
direct photon production presents a favorable 
opportunity to test the process dependence of the Sivers function, or the effect of the associated ISIs. 

\section{Summary}
\label{sum}
In this paper, we have studied the single transverse spin asymmetries in the single inclusive particle production in hadronic collisions. We point out the Sivers functions in such processes are generally different from those probed in the SIDIS process because of different initial- and final-state interactions. By carefully taking into account the process-dependence in the Sivers functions (under one-gluon exchange approximation), we derive a new formalism within the framework of GPM approach. We find this formalism has close connections with the collinear twist-3 approach.
With our modified GPM formalism, we make predictions for the inclusive $\pi^0$ and direct photon production in $pp$ collisions at RHIC energies. We find that the asymmetries predicted from the modified GPM formalism are opposite to those in the conventional GPM approach. This sign difference comes from the color gauge interaction, which has the same origin as the sign change for Sivers functions between SIDIS and DY processes. Our predictions about the sign are consistent with those from the twist-3 collinear factorization approach. We encourage a global analysis of both SIDIS and $pp$ experimental data using  this  modified GPM formalism.

\section*{Acknowledgments}

We are grateful to M.~Anselmino, U.~D'Alesio, A.~Metz, P.~Mulders, F.~Murgia,  J.~W.~Qiu, W.~Vogelsang, F.~Yuan and J.~Zhou for useful discussions and comments. L.G. acknowledges support from U.S. Department of Energy under contract DE-FG02-07ER41460. Z.K. is grateful to RIKEN, Brookhaven National Laboratory, and the U.S. Department of Energy (Contract No.~DE-AC02-98CH10886) for supporting this work.


\begin{thebibliography}{99}

\bibitem{D'Alesio:2007jt}
  For reviews, see:
  U.~D'Alesio and F.~Murgia,
  %``Azimuthal and Single Spin Asymmetries in Hard Scattering Processes,''
  Prog.\ Part.\ Nucl.\ Phys.\  {\bf 61}, 394 (2008)
  [arXiv:0712.4328 [hep-ph]].
  
 \bibitem{HERMES}
  A.~Airapetian {\it et al.}  [HERMES Collaboration],
  %``Single-spin asymmetries in semi-inclusive deep-inelastic scattering on  a
  %transversely polarized hydrogen target,''
  Phys.\ Rev.\ Lett.\  {\bf 94}, 012002 (2005)
  [arXiv:hep-ex/0408013];
  %``Observation of the Naive-T-odd Sivers Effect in Deep-Inelastic
  %Scattering,''
  Phys.\ Rev.\ Lett.\  {\bf 103}, 152002 (2009)
  [arXiv:0906.3918 [hep-ex]].
  
\bibitem{COMPASS}
  V.~Y.~Alexakhin {\it et al.}  [COMPASS Collaboration],
  %``First measurement of the transverse spin asymmetries of the deuteron in
  %semi-inclusive deep inelastic scattering,''
  Phys.\ Rev.\ Lett.\  {\bf 94}, 202002 (2005)
  [arXiv:hep-ex/0503002];
  A.~Martin  [COMPASS Collaboration],
  %``COMPASS Results on Transverse Single-Spin Asymmetries,''
  Czech.\ J.\ Phys.\  {\bf 56}, F33 (2006)
  [arXiv:hep-ex/0702002];
  M.~Alekseev {\it et al.}  [COMPASS Collaboration],
  %``Collins and Sivers asymmetries for pions and kaons in muon-deuteron DIS,''
  Phys.\ Lett.\  B {\bf 673}, 127 (2009)
  [arXiv:0802.2160 [hep-ex]].

\bibitem{SSA-rhic}
  J.~Adams {\it et al.}  [STAR Collaboration],
  %``Cross sections and transverse single-spin asymmetries in forward  neutral
  %pion production from proton collisions at s**(1/2) = 200-GeV,''
  Phys.\ Rev.\ Lett.\  {\bf 92}, 171801 (2004)
  [arXiv:hep-ex/0310058];
B.~I.~Abelev {\it et al.}  [STAR Collaboration],
%``Measurement of Transverse Single-Spin Asymmetries for Di-Jet Production
%in Proton-Proton Collisions at $\sqrt{s} = 200$ GeV,''
  Phys.\ Rev.\ Lett.\  {\bf 99}, 142003 (2007)
  [arXiv:0705.4629 [hep-ex]];
  %``Forward Neutral Pion Transverse Single Spin Asymmetries in p+p Collisions
  %at \sqrt{s}=200 GeV,''
  Phys.\ Rev.\ Lett.\  {\bf 101}, 222001 (2008)
  [arXiv:0801.2990 [hep-ex]];
  S.~S.~Adler {\it et al.}  [PHENIX Collaboration],
  %``Measurement of transverse single-spin asymmetries for mid-rapidity
  %production of neutral pions and charged hadrons in polarized p + p
  %collisions at s**(1/2) = 200-GeV,''
  Phys.\ Rev.\ Lett.\  {\bf 95}, 202001 (2005)
  [arXiv:hep-ex/0507073];
  I.~Arsene {\it et al.}  [BRAHMS Collaboration],
  %``Single Transverse Spin Asymmetries of Identified Charged Hadrons in
  %Polarized p+p Collisions at $\sqrt{s}$ = 62.4 GeV,''
Phys.\ Rev.\ Lett.\  {\bf 101}, 042001 (2008)
  [arXiv:0801.1078 [nucl-ex]].

\bibitem{Brodsky:2002cx}
  S.~J.~Brodsky, D.~S.~Hwang and I.~Schmidt,
  %``Final-state interactions and single-spin asymmetries in semi-inclusive
  %deep inelastic scattering,''
  Phys.\ Lett.\  B {\bf 530}, 99 (2002)
  [arXiv:hep-ph/0201296];
  S.~J.~Brodsky, D.~S.~Hwang and I.~Schmidt,
  %``Initial-state interactions and single-spin asymmetries in Drell-Yan
  %processes,''
  Nucl.\ Phys.\  B {\bf 642}, 344 (2002)
  [arXiv:hep-ph/0206259].

\bibitem{Collins:2002kn}
  J.~C.~Collins,
  %``Leading-twist Single-transverse-spin asymmetries: Drell-Yan and
  %Deep-Inelastic Scattering,''
  Phys.\ Lett.\  B {\bf 536}, 43 (2002)
  [arXiv:hep-ph/0204004].

\bibitem{TMD-dis}
X.~d.~Ji and F.~Yuan,
  %``Parton distributions in light-cone gauge: Where are the final-state
  %interactions?,''
  Phys.\ Lett.\  B {\bf 543}, 66 (2002)
  [arXiv:hep-ph/0206057];
A.~V.~Belitsky, X.~Ji and F.~Yuan,
  %``Final state interactions and gauge invariant parton distributions,''
  Nucl.\ Phys.\  B {\bf 656}, 165 (2003)
  [arXiv:hep-ph/0208038].

\bibitem{boermulders} 
D.~Boer, P.~J.~Mulders and F.~Pijlman,
  %``Universality of T-odd effects in single spin and azimuthal asymmetries,''
  Nucl.\ Phys.\  B {\bf 667}, 201 (2003)
  [arXiv:hep-ph/0303034].

\bibitem{Siv90}
D.~W.~Sivers,
%``Single Spin Production Asymmetries From The Hard Scattering Of
% Point- Like
%Constituents,''
Phys.\ Rev.\ D {\bf 41}, 83 (1990);
%``Hard Scattering Scaling Laws For Single Spin Production
% Asymmetries,''
Phys.\ Rev.\ D {\bf 43}, 261 (1991).

\bibitem{mulders}  
A.~Bacchetta, C.~J.~Bomhof, P.~J.~Mulders and F.~Pijlman,
  %``Single spin asymmetries in hadron hadron collisions,''
  Phys.\ Rev.\  D {\bf 72}, 034030 (2005)
  [arXiv:hep-ph/0505268];
C.~J.~Bomhof, P.~J.~Mulders and F.~Pijlman,
  %``The construction of gauge-links in arbitrary hard processes,''
  Eur.\ Phys.\ J.\  C {\bf 47}, 147 (2006)
  [arXiv:hep-ph/0601171].

\bibitem{Kang:2009bp}
  Z.~B.~Kang and J.~W.~Qiu,
  %``Testing the Time-Reversal Modified Universality of the Sivers Function,''
  Phys.\ Rev.\ Lett.\  {\bf 103}, 172001 (2009)
  [arXiv:0903.3629 [hep-ph]].
  
\bibitem{SSA-fixed-tgt}
  G.~Bunce {\it et al.},
  %``Lambda0 Hyperon Polarization In Inclusive Production By 300-Gev Protons On
  %Beryllium,''
  Phys.\ Rev.\ Lett.\  {\bf 36}, 1113 (1976);
  D.~L.~Adams {\it et al.}  [E581 and E704 Collaborations],
  %``Comparison of spin asymmetries and cross-sections in pi0
  % production by 200-GeV polarized anti-protons and protons,''
  Phys.\ Lett.\ B {\bf 261}, 201 (1991);
  D.~L.~Adams {\it et al.}  [FNAL-E704 Collaboration],
  %``Analyzing power in inclusive pi+ and pi- production at high x(F)
  % with a 200-GeV polarized proton beam,''
  Phys.\ Lett.\ B {\bf 264}, 462 (1991);
  K.~Krueger {\it et al.}, Phys.\ Lett.\ B {\bf 459}, 412 (1999).

\bibitem{Efremov}
  A.~V.~Efremov and O.~V.~Teryaev,
  %``On Spin Effects In Quantum Chromodynamics,''
  Sov.\ J.\ Nucl.\ Phys.\  {\bf 36}, 140 (1982)
  [Yad.\ Fiz.\  {\bf 36}, 242 (1982)];
  %``The Transversal Polarization In Quantum Chromodynamics,''
  Sov.\ J.\ Nucl.\ Phys.\  {\bf 39}, 962 (1984)
  [Yad.\ Fiz.\  {\bf 39}, 1517 (1984)];
  %``QCD Asymmetry And Polarized Hadron Structure Functions,''
  Phys.\ Lett.\ B {\bf 150}, 383 (1985).

\bibitem{qiu-sterman}
  J.~W.~Qiu and G.~Sterman,
  %``Single Transverse Spin Asymmetries,''
  Phys.\ Rev.\ Lett.\  {\bf 67}, 2264 (1991);
  %``Single transverse spin asymmetries in direct photon production,''
  Nucl.\ Phys.\  B {\bf 378}, 52 (1992);
%``Single transverse-spin asymmetries in hadronic pion production,''
Phys.\ Rev.\ D {\bf 59}, 014004 (1999).

\bibitem{inclupi}
  C.~Kouvaris, J.~W.~Qiu, W.~Vogelsang and F.~Yuan,
  %``Single transverse-spin asymmetry in high transverse momentum pion
  %production in p p collisions,''
  Phys.\ Rev.\  D {\bf 74}, 114013 (2006)
  [arXiv:hep-ph/0609238].
  
\bibitem{applytwist}
  Z.~B.~Kang, J.~W.~Qiu, W.~Vogelsang and F.~Yuan,
  %``Accessing tri-gluon correlations in the nucleon via the single spin
  %asymmetry in open charm production,''
  Phys.\ Rev.\  D {\bf 78}, 114013 (2008)
  [arXiv:0810.3333 [hep-ph]];
  Z.~B.~Kang, F.~Yuan and J.~Zhou,
  %``Twist-three Fragmentation Function Contribution to the Single Spin
  %Asymmetry in pp Collisions,''
  Phys.\ Lett.\  B {\bf 691}, 243 (2010)
  [arXiv:1002.0399 [hep-ph]].

\bibitem{Anselmino}
M.~Anselmino, M.~Boglione and F.~Murgia,
  %``Single spin asymmetry for p (polarized) p $\to$ pi X in perturbative QCD,''
  Phys.\ Lett.\  B {\bf 362}, 164 (1995)
  [arXiv:hep-ph/9503290];
  M.~Anselmino and F.~Murgia,
  %``Single spin asymmetries in p(pol.) p and anti-p(pol.) p inclusive
  %processes,''
  Phys.\ Lett.\  B {\bf 442}, 470 (1998)
  [arXiv:hep-ph/9808426];
  U.~D'Alesio and F.~Murgia,
  %``Parton intrinsic motion in inclusive particle production: Unpolarized
  %cross sections, single spin asymmetries and the Sivers effect,''
  Phys.\ Rev.\  D {\bf 70}, 074009 (2004)
  [arXiv:hep-ph/0408092];
  M.~Anselmino, M.~Boglione, U.~D'Alesio, E.~Leader and F.~Murgia,
  %``Parton intrinsic motion: Suppression of the Collins mechanism for
  %transverse single spin asymmetries in p(pol.) p --> pi X,''
  Phys.\ Rev.\  D {\bf 71}, 014002 (2005)
  [arXiv:hep-ph/0408356];
  M.~Anselmino, M.~Boglione, U.~D'Alesio, E.~Leader, S.~Melis and F.~Murgia,
  %``The general partonic structure for hadronic spin asymmetries,''
  Phys.\ Rev.\  D {\bf 73}, 014020 (2006)
  [arXiv:hep-ph/0509035].

\bibitem{Boglione:2007dm}
  M.~Boglione, U.~D'Alesio and F.~Murgia,
  %``Single spin asymmetries in inclusive hadron production from SIDIS to
  %hadronic collisions: universality and phenomenology,''
  Phys.\ Rev.\  D {\bf 77}, 051502 (2008)
  [arXiv:0712.4240 [hep-ph]].
   
\bibitem{Anselmino:2008uy}
  M.~Anselmino, M.~Boglione, U.~D'Alesio, E.~Leader, S.~Melis and F.~Murgia,
  %``Sivers and Collins effects in polarized pp scattering processes,''
  arXiv:0809.3743 [hep-ph].

\bibitem{Bacchetta:2004jz}
  A.~Bacchetta, U.~D'Alesio, M.~Diehl and C.~A.~Miller,
  %``Single-spin asymmetries: The Trento conventions,''
  Phys.\ Rev.\  D {\bf 70}, 117504 (2004)
  [arXiv:hep-ph/0410050].

\bibitem{oldsiv}
  M.~Anselmino, M.~Boglione, U.~D'Alesio, A.~Kotzinian, F.~Murgia and A.~Prokudin,
  %``Extracting the Sivers function from polarized SIDIS data and making
  %predictions,''
  Phys.\ Rev.\  D {\bf 72}, 094007 (2005)
  [Erratum-ibid.\  D {\bf 72}, 099903 (2005)]
  [arXiv:hep-ph/0507181].
  
\bibitem{newsiv}
  M.~Anselmino {\it et al.},
  %``Sivers Effect for Pion and Kaon Production in Semi-Inclusive Deep Inelastic
  %Scattering,''
  Eur.\ Phys.\ J.\  A {\bf 39}, 89 (2009)
  [arXiv:0805.2677 [hep-ph]].
    
\bibitem{Field:1976ve}
  R.~D.~Field and R.~P.~Feynman,
  %``Quark Elastic Scattering As A Source Of High Transverse Momentum Mesons,''
  Phys.\ Rev.\  D {\bf 15}, 2590 (1977);
  R.~P.~Feynman, R.~D.~Field and G.~C.~Fox,
  %``Quantum-chromodynamic approach for the large-transverse-momentum
  %production of particles and jets,''
  Phys.\ Rev.\  D {\bf 18}, 3320 (1978).

\bibitem{Collins:1981uk}
  J.~C.~Collins and D.~E.~Soper,
  %``Back-To-Back Jets In QCD,''
  Nucl.\ Phys.\  B {\bf 193}, 381 (1981)
  [Erratum-ibid.\  B {\bf 213}, 545 (1983)];
  J.~C.~Collins, D.~E.~Soper and G.~F.~Sterman,
  %``Transverse Momentum Distribution In Drell-Yan Pair And W And Z Boson
  %Production,''
  Nucl.\ Phys.\  B {\bf 250}, 199 (1985).

\bibitem{Ji:2004wu}
  X.~d.~Ji, J.~p.~Ma and F.~Yuan,
  %``QCD factorization for semi-inclusive deep-inelastic scattering at low
  %transverse momentum,''
  Phys.\ Rev.\  D {\bf 71}, 034005 (2005)
  [arXiv:hep-ph/0404183].
  
\bibitem{Qiu:2007ey}
  J.~W.~Qiu, W.~Vogelsang and F.~Yuan,
  %``Single Transverse-Spin Asymmetry in Hadronic Dijet Production,''
  Phys.\ Rev.\  D {\bf 76}, 074029 (2007)
  [arXiv:0706.1196 [hep-ph]].

\bibitem{Yuan:2008vn}
  F.~Yuan,
  %``Heavy Quarkonium Production in Single Transverse Polarized High Energy
  %Scattering,''
  Phys.\ Rev.\  D {\bf 78}, 014024 (2008)
  [arXiv:0801.4357 [hep-ph]].

\bibitem{tmdbreak}
  J.~Collins and J.~W.~Qiu,
  %``$k_{T}$ factorization is violated in production of high-transverse-momentum
  %particles in hadron-hadron collisions,''
  Phys.\ Rev.\  D {\bf 75}, 114014 (2007)
  [arXiv:0705.2141 [hep-ph]];
  J.~Collins,
  %``2-soft-gluon exchange and factorization breaking,''
  arXiv:0708.4410 [hep-ph];
  W.~Vogelsang and F.~Yuan,
  %``Hadronic Dijet Imbalance and Transverse-Momentum Dependent Parton
  %Distributions,''
  Phys.\ Rev.\  D {\bf 76}, 094013 (2007)
  [arXiv:0708.4398 [hep-ph]];
  T.~C.~Rogers and P.~J.~Mulders,
  %``No Generalized TMD-Factorization in the Hadro-Production of High Transverse
  %Momentum Hadrons,''
  Phys.\ Rev.\  D {\bf 81}, 094006 (2010)
  [arXiv:1001.2977 [hep-ph]].

\bibitem{GK}
L.~Gamberg and Z.~B.~Kang, in preparation.

\bibitem{SFP}
  Y.~Koike and T.~Tomita,
  %``Soft-fermion-pole contribution to single-spin asymmetry for pion production
  %in pp collisions,''
  Phys.\ Lett.\  B {\bf 675}, 181 (2009)
  [arXiv:0903.1923 [hep-ph]];
  K.~Kanazawa and Y.~Koike,
  %``New Analysis of the Single Transverse-Spin Asymmetry for Hadron Production
  %at RHIC,''
  Phys.\ Rev.\  D {\bf 82}, 034009 (2010)
  [arXiv:1005.1468 [hep-ph]];
  Z.~B.~Kang, J.~W.~Qiu and H.~Zhang,
  %``Quark-gluon correlation functions relevant to single transverse spin
  %asymmetries,''
  Phys.\ Rev.\  D {\bf 81}, 114030 (2010)
  [arXiv:1004.4183 [hep-ph]].

\bibitem{Qiu:1990cu}
  J.~W.~Qiu and G.~Sterman,
  %``High twist effects in hadronic collisions,''
  AIP Conf.\ Proc.\  {\bf 223}, 249 (1991).
  
\bibitem{Gluck:1998xa}
  M.~Gluck, E.~Reya and A.~Vogt,
  %``Dynamical parton distributions revisited,''
  Eur.\ Phys.\ J.\  C {\bf 5}, 461 (1998)
  [arXiv:hep-ph/9806404].

\bibitem{Owens:1986mp}
  J.~F.~Owens,
  %``Large Momentum Transfer Production Of Direct Photons, Jets, And
  %Particles,''
  Rev.\ Mod.\ Phys.\  {\bf 59}, 465 (1987).
  
\bibitem{Kang:2010vd}
  Z.~B.~Kang and F.~Yuan,
  %``Dihadron Azimuthal Correlation from Collins Effect in Unpolarized Hadron
  %Collisions,''
  Phys.\ Rev.\  D {\bf 81}, 054007 (2010)
  [arXiv:1001.0247 [hep-ph]].

\bibitem{deFlorian:2007aj}
  D.~de Florian, R.~Sassot and M.~Stratmann,
  %``Global analysis of fragmentation functions for pions and kaons and their
  %uncertainties,''
  Phys.\ Rev.\  D {\bf 75}, 114010 (2007)
  [arXiv:hep-ph/0703242].

\bibitem{Bacchetta:2007sz}
  A.~Bacchetta, C.~Bomhof, U.~D'Alesio, P.~J.~Mulders and F.~Murgia,
  %``The Sivers single-spin asymmetry in photon - jet production,''
  Phys.\ Rev.\ Lett.\  {\bf 99}, 212002 (2007)
  [arXiv:hep-ph/0703153].
              
\bibitem{Collins:1992kk}
  J.~C.~Collins,
  %``Fragmentation of transversely polarized quarks probed in transverse
  %momentum distributions,''
  Nucl.\ Phys.\  B {\bf 396}, 161 (1993)
  [arXiv:hep-ph/9208213].


\end{thebibliography}
\end{document}